\documentclass[11pt]{article}
\usepackage{amsfonts,amssymb}
\usepackage{color}
\usepackage{pstricks,pst-node}
\usepackage{pst-plot}
\usepackage{mathtools}
\catcode`\@=11
\def\marginnote#1{}

\newcount\hour
\newcount\minute
\newtoks\amorpm
\hour=\time\divide\hour by60 \minute=\time{\multiply\hour by60
\global\advance\minute by-\hour}
\edef\standardtime{{\ifnum\hour<12 \global\amorpm={am}%
        \else\global\amorpm={pm}\advance\hour by-12 \fi
        \ifnum\hour=0 \hour=12 \fi
        \number\hour:\ifnum\minute<10 0\fi\number\minute\the\amorpm}}
\edef\militarytime{\number\hour:\ifnum\minute<10 0\fi\number\minute}

\def\appendix#1{
\addtocounter{section}{1} \setcounter{equation}{0}
\renewcommand{\thesection}{\Alph{section}}
\section*{Appendix \thesection\protect\indent\quad
#1}
}
\renewcommand{\theequation}{\thesection.\arabic{equation}}

\def\draftlabel#1{{\@bsphack\if@filesw {\let\thepage\relax
      \xdef\@gtempa{\write\@auxout{\string
          \newlabel{#1}{{\@currentlabel}{\thepage}}}}}\@gtempa \if@nobreak
    \ifvmode\nobreak\fi\fi\fi\@esphack} \gdef\@eqnlabel{#1}}
    \def\@eqnlabel{}
\def\@vacuum{}
\def\draftmarginnote#1{\marginpar{\raggedright\scriptsize\tt#1}}

\def\draft{
  \oddsidemargin -.5truein
  \def\@oddfoot{\footnotesize \sl preliminary draft \hfil
    \rm\thepage\hfil\sl\today\quad\militarytime}
  \let\@evenfoot\@oddfoot \overfullrule 3pt
    \let\label=\draftlabel
    \let\marginnote=\draftmarginnote
  \def\@eqnnum{(\theequation)\rlap{\kern\marginparsep\tt\@eqnlabel}%
    \global\let\@eqnlabel\@vacuum}

  }

\voffset= - 1.0in \hoffset= - 1.0in

\def\be{\begin{equation}}
\def\ee{\end{equation}}
\def\bea{\begin{eqnarray}}
\def\eea{\end{eqnarray}}
\def\<{\langle}
\def\>{\rangle}

\def\Im{{\rm Im}}

\def\tr{{\mathrm{tr\,}}}

\def\1N{${\cal N}=1$}
\def\4N{${\cal N}=4$}

\def\bea{\begin{eqnarray}}
\def\eea{\end{eqnarray}}

\def\beq{\begin{equation}}
\def\eeq{\end{equation}}
\def\ba{\beq\begin{array}{c}}
\def\ea{\end{array}\eeq}

\@ifundefined{theorem@style}{\RequirePackage{theorem}}{}

\gdef\th@plain{\normalfont\slshape
  \def\@begintheorem##1##2{%
\item[\hskip\parindent\hskip\labelsep\theorem@headerfont ##1\ ##2\unskip.]}%
\def\@opargbegintheorem##1##2##3{%
\item[\hskip\parindent
\ifx\empty##1\else\hskip\labelsep\fi\theorem@headerfont ##1\ ##2\unskip]{\theorem@headerfont{\rm ##3}.} }}

\gdef\th@definition{\normalfont
  \def\@begintheorem##1##2{%
\item[\hskip\parindent\hskip\labelsep\theorem@headerfont ##1\ ##2\unskip.]}%
\def\@opargbegintheorem##1##2##3{%
\item[\hskip\parindent
\ifx\empty##1\else\hskip\labelsep\fi\theorem@headerfont ##1\ ##2\unskip]{\theorem@headerfont{\rm ##3}.} }}

\theoremstyle{plain}
\newtheorem{theorem}{Theorem}
\newtheorem{lemma}{Lemma}
\newtheorem{corollary}{Corollary}
\newtheorem{proposition}{Proposition}

\theoremstyle{definition}

\let\text=\mathrm

\def\beq{\begin{equation}}
\def\eeq{\end{equation}}
\def\bea{\begin{eqnarray}}
\def\eea{\end{eqnarray}}

\newcommand{\cpict}[3]{
\dimen1=#1\advance\dimen1 by-\hsize\divide\dimen1 by-2 \vtop to #2{
\noindent\hskip\dimen1{\special{em:graph #3.bmp}} \vfil}\hskip-2cm }

\let\@@savethanks\thanks
\def\thanks#1{\gdef\thefootnote{\alph{footnote}}\@@savethanks{#1}}

\makeatletter
\def\blfootnote{\xdef\@thefnmark{}\@footnotetext}
\makeatother

\begin{document}

\title{The Harer--Zagier recursion for an irregular spectral curve}

\author{Leonid O. Chekhov\thanks{Steklov Mathematical Institute, Moscow, Russia. The work was funded by grant from the Russian Science Foundation (Project No. 14-50-00005).}}

\date{}


\maketitle

\begin{abstract}
We derive the Do and Norbury recursion formula for the one-loop mean of an irregular spectral curve from
a variant of replica method by Brez\'in and Hikami. We express this recursion in special times in which all
terms $W_1^{(g)}$ of the genus expansion of the one-loop mean are polynomials. We generalize this recursion to the case of generalized Laguerre ensemble.
\end{abstract}

\section{Introduction}\label{s:i}

Multi-trace means $\<\prod_{i=1}^s \tr H^{k_i}\>^{{\mathrm{conn}}}$ taken over ensembles of classical orthogonal polynomials (Hermite, Laguerre and Legendre) contain complete information about free energies of the corresponding matrix models with arbitrary (polynomial) potentials $V(H)=\sum_k \frac 1k r_k \tr H^k$ and with possible restrictions on the admissible domain of eigenvalue distribution. In the Hermitian case, we have $r_k=t_k-\delta_{k,2}$ and we consider a perturbative expansion in times $t_k$ averaged with the Gaussian measure over the whole real axis. In the Laguerre (or Poisson) case under consideration, we have $r_k=t_k-\delta_{k,1}$ and require all eigenvalues of $H$ to be nonnegative. Finally, in the Legendre case, $r_k=t_k$ and we restrict eigenvalues to a finite interval (commonly, $[-2,2]$). A general tool for a systematic evaluation of multi-trace mean generating functions called multi-loop means,
\beq
W_s(x_1,\dots,x_s):=\Bigl\<\prod_{i=1}^s \Bigl(\tr\frac{1}{x_i-H}\Bigr) \Bigr\>^{\text{conn}},
\eeq
is the topological recursion method \cite{Ey,ChEy,CEO,EOrinv,AlMM}. Multi-loop means always admit a $1/N$-expansion (where $N$ is the size of matrices),
\beq
W_s(x_1,\dots,x_s)=\sum_{g=0}^\infty N^{2-2g-s} W^{(g)}_s(x_1,\dots,x_s),
\eeq
and using the topological recursion we construct all terms $W^{(g)}_s$ out of $W^{(0)}_1(x)=y(x)+V'(x)/2$ and $W^{(0)}_2(x_1,x_2)$, the latter is always a universal object: $W^{(0)}_2(x_1,x_2)dx_1dx_2=\frac 12 \Bigl(B(x_1,x_2)- \frac{dx_1\,dx_2}{(x_1-x_2)^2}\Bigr)$, where $B(x_1,x_2)$ is a (canonically normalized) Bergmann bi-differential depending only on the spectral curve $\Sigma(x,y)=0$ of the model (here $y$ is related to $W_1^{(0)}(x)$ as above). The topological recursion method determines all $W_s^{(g)}$ in a canonical way but it involves tedious calculations and, presumably, we cannot go beyond $g+s\sim 10{-}15$ even with the help of supercomputers. There are however more economical tools for calculating one-point correlation functions $W_1^{(g)}(x)$. In this case, the celebrated three-term recursion relation found by Harer and Zagier \cite{HZ} for Gaussian ensembles enables finding $W^{(g)}_1(x)$ for incredibly high $g$. This recursion was improved in \cite{ACNP,HT}  where it was shown that at any genus $g>0$ we have the polynomial representation 
\beq
W_1^{(g)}(e^\lambda+e^{-\lambda})=\sum_{k=1}^{g-1} b_k^{(g)}\frac{1}{(e^\lambda-e^{-\lambda})^{4g+2k+1}},\quad g>0
\label{WGauss}
\eeq
and all coefficients $b_k^{(g)}$ are positive integers enjoying their own three-term recursion relation
\beq
\label{recurrence}
(4g+2k+6)b^{(g+1)}_k=(4g+2k+1)(4g+2k+3)\Bigl[(4g+2k+2)b^{(g)}_k+4(4g+2k-1)b^{(g)}_{k-1}\Bigr].
\eeq

It is well known (see, e.g., \cite{AJM}) that the Laguerre, or Poisson ensemble under consideration here is in bijection with the complex matrix model, or with the bicolored maps. For these bicolored maps, Bernardi and Chapuy \cite{BC} (see also \cite{CFF}) derived  a combinatorial formula describing a number of unicellular maps of a given genus (see also \cite{Jackson88} for the early accounting of these maps). Using these results,  Do and Norbury found recently (Theorem 4.1 in \cite{DN}) the three-term recursion relation for the one-loop correlation functions of an irregular spectral curve related to the Laguerre ensemble.  In the present note, we derive this three-term relation using an adaptation of the Brez{\'i}n--Hikami replica method \cite{BreHik} (see \cite{MS10} for development of this method as regarding multi-loop correlation functions). This allows obtaining expressions in terms of $s$-fold integrals for bicolored maps with $s$ cells, or faces. As an application of this technique, we derive the Harer--Zagier--Do--Norbury recursion formula using only manipulations with the corresponding integral (\ref{rep-N}). We then express $W_1^{(g)}$ in a form similar to (\ref{WGauss}) in terms of special times $v_k(\lambda)$, which satisfy their own recursion relations described in Lemma~\ref{lem:vk}.

We then show that using a variant of the same replica method we can derive recursion formulas for one-loop means in the case of generalized Laguerre ensemble with the measure of integration $x^k e^{-Nx}$ on $[0,\infty)$. We demonstrate that the correlation function $\Bigl\langle \tr e^{-uH} \Bigr\rangle$ averaged with this measure satisfies a linear differential equation of order $2k$ (for $k\in {\mathbb Z}_+$) and we describe a regular method of deriving this equation for any $k$. We then derive the corresponding equations in the cases $k=1$ and $k=2$. Worth mentioning is that, although theories with fixed $k$ lose some nice properties of theories with $k=\alpha N$ for a fixed $\alpha$: their $1/N$-expansion contains odd powers of $1/N$ and their partition functions presumably are not tau functions of  integrable hierarchies (at least, for arbitrary $k$), nevertheless, it was shown in \cite{Alexandrov1,Alexandrov2} that generalizing the Kontsevich matrix model by adding a logarithmic term in the potential (which corresponds to the case $k=1$) yields a generating function for intersection indices of the open moduli spaces (see \cite{Pandharipande,Buryak1,Buryak2}). A geometrical interpretation of models with $k>1$ remains open as yet.  

\section{Bicolored maps and the Laguerre ensembles}\label{s:B2L}
We begin with a brief reminder on the relation between bicolored maps and ensembles of Hermitian matrices with restrictions on eigenvalue supports. We relate this problem to that of constructing  generating functions for branched coverings $C_g\to CP_1$. It was observed in \cite{Zograf}, \cite{AmCh14} that these generating functions in the case of Grothendieck's \emph{dessins d'enfant} where we have exactly three branching points on the projective line are related to complex matrix models. Generalizations to cases of higher (fixed) numbers of branching points use multi-matrix models \cite{AmCh15}, which were previously studied in \cite{AIK}. The spectral curve and the whole topological recursion procedure for these models were constructed in \cite{AmCh15}. We refer the reader to these papers for details. Here we mention that (the derivatives of) the generating function for Grothendieck's dessins d'enfant with three branching points in the case where we assume that the profile at one of these points (say, infinity) is given by a Young diagram $\{\mu\}=(\mu_1\ge \mu_2\ge
\cdots \ge \mu_s>0)$ and we account for the numbers $q_1$ and $q_2$ of preimages of two other branching points is the integral
\beq
\frac{1}{\mathcal Z}\int DB\,DB^\dagger\,e^{-N\tr B B^\dagger} \Bigl[\prod_{i=1}^s \frac{1}{\mu_i}\tr\bigl[(BB^\dagger)^{\mu_i}\bigr]\Bigr]^{\text{conn}}
=\sum_{g=0}^\infty \sum_{q_1,q_2} N^{2-2g} \gamma_1^{q_1}\gamma_2^{q_2} H_{\{\mu\},t_1,t_2},
\label{BB-1}
\eeq
where we integrate with the standard Haar measure $\prod_{i,j} d\Re B_{i,j}d\Im B_{i,j}$ over \emph{rectangular} $(\gamma_1N \times \gamma_2 N)$
complex matrices. Note that generating functions for (\ref{BB-1}) with the fixed $s$ are multi-loop means
\beq
\frac{1}{\mathcal Z}\int DB\,DB^\dagger\,e^{-N\tr B B^\dagger} \Bigl[\prod_{i=1}^s \tr \frac{1}{x_i-BB^\dagger}\Bigr]^{\text{conn}}.
\label{BB-2}
\eeq
We can further transform (\ref{BB-2}) using the Marchenko--Pastur law \cite{MP}: 
for $\gamma_2 \ge \gamma_1$ we can replace integration w.r.t. complex matrices
$B,B^\dagger$ by that over the Hermitian $(\gamma_1\times \gamma_1)$-matrix $H=B^\dagger B$, which has to be positive-definite:
$$
\frac{1}{\mathcal Z}\int DB\,DB^\dagger\,e^{-N\tr B B^\dagger} \Bigl[\prod_{i=1}^s \tr \frac{1}{x_i-BB^\dagger}\Bigr]^{\text{conn}}
=\frac{1}{\mathcal Z}\int DH_{\ge0}\,e^{-N\tr H} [\det H]^{(\gamma_2-\gamma_1)N}
\Bigl[\prod_{i=1}^s \tr \frac{1}{x_i-H}\Bigr]^{\text{conn}}.
$$
Meanwhile, all corresponding partition functions are $\tau$-functions of the Kadomtsev--Petviashvili hierarchy, as was shown in \cite{HO,Kazarian,AMMN,KZ,HP}.

\section{The replica method for the Laguerre ensemble}\label{s:replica}

Loop means $W_s(x_1,\dots,x_s)$ of the Laguerre, or Poisson, ensemble are given by $N$-fold integrals over eigenvalues $\lambda_i$ of a positive-definite Hermitian matrix $H$:
\bea
W_s(x_1,\dots,x_s)&:=&\frac1{\mathcal Z}\int\cdots\int_0^\infty d\lambda_1\cdots d\lambda_N e^{-N\sum_{i=1}^N \lambda_i}\Delta^2(\lambda)
\prod_{k=1}^s\Bigl(\sum_{i=1}^N \frac1{x_k-\lambda_i}\Bigr)^{\text{conn}}\nonumber\\
&:=&\Bigl\langle  \prod_{k=1}^s \tr \frac{1}{x_k-H}\Bigr\rangle^{\text{conn}}
\label{loop-s}
\eea
where $\Delta(\lambda)=\prod_{i<j}(\lambda_i-\lambda_j)$ is the Vandermonde determinant and we select the connected part of the correlation function in a standard way (for $s=1$ it is $W_1(x)$ itself, $W_2(x_1,x_2)=W(x_1,x_2)^{\text{total}}-W_1(x_1)W_1(x_2)$, etc.). For the presentation brevity, we denote the sums over $i$ by the trace symbol.  

It is technically simpler to start with integrals of exponentials, 
\beq
\Bigl\langle  \prod_{k=1}^s \tr e^{u_k H}\Bigr\rangle.
\label{exp-u}
\eeq
In order to find an $s$-fold integral representation of the above correlation functions, we use the replica method of \cite{BreHik}. It is based on the following trick: under the integration sign in (\ref{exp-u}) we insert the following integral over unitary $N\times N$-matrix $U$ depending on additional $N$ external variables $a_i$, $i=1,\dots,N$:
$$
\frac 1{\hbox{Vol\,} U}\int dU \,e^{\sum_{i,j}U_{ij}\lambda_j U_{ji}^\ast a_i}=\frac{\det_{i,j} e^{\lambda_j a_i}}{\Delta(\lambda)\Delta(a)}.
$$
We have used here the celebrated Charish-Chandra--Itzykson--Zuber integration formula. Note that when setting all $a_i=0$, this integral just becomes the unity.

Because the integrand in (\ref{loop-s}) is totally symmetric w.r.t. all $\lambda_j$ we can replace the determinant  of $e^{\lambda_j a_i}$ by the product $\prod_{i=1}^N e^{a_i\lambda_i}$. We now have to evaluate the integral
\bea
&{}&\sum_{\{k_i\}_{i{=}1}^s} \frac1{\mathcal Z}\int_0^\infty\cdots\int_0^\infty d\lambda_1\cdots d\lambda_N e^{-N\sum_{i=1}^N \lambda_i(1-a_i-\sum_{j=1}^s \delta_{i,k_j}u_j )}\frac{\Delta(\lambda)}{\Delta(a)}\nonumber\\
&{}&=\sum_{\{k_i\}_{i{=}1}^s} \prod_{i=1}^N \frac 1{(1-a_i-\sum_{j=1}^s \delta_{i,k_j}u_j )^N}\frac{\Delta(1-a_i-\sum_{j=1}^s \delta_{i,k_j}u_j )}{\Delta(a)}.
\label{BH-1}
\eea
We now represent the result of integration in (\ref{BH-1}) as an $s$-fold contour integral. For this, we assume that the contour $C_0$ encircles all $a_i$ but not any other possible singularity of an integrand. We then use that in the ratio of the Vandermonde determinants (recalling that $\Delta(f)=\prod_{i<j}(f_i-f_j)$) only those terms for which $f_i$ or/and $f_j$ are different contribute. 

We begin with the case $s=1$ where we have
$$
\frac{\Delta(1-a_i-u \delta_{i,k} )}{\Delta(a)}=\prod_{j\ne k} \frac{a_k+u-a_j}{a_k-a_j}, \quad (s=1)
$$   
and, analogously,
$$
\prod_{i=1}^N \frac 1{(1-a_i-u \delta_{i,k} )^N}=\prod_{i=1}^N \frac 1{(1-a_i)^N} \frac{(1-a_k)^N}{(1-a_k-u)^N}, \quad (s=1).
$$
Now the crucial step follows. We can present the sum in (\ref{BH-1}) as a sum of residues at the points $z=a_i$ of the integral,
\beq
\Bigl\langle \tr e^{uH} \Bigr\rangle_a =\prod_{i=1}^N \frac 1{(1-a_i)^N}\frac 1u \frac 1{2\pi i} \oint_{C_0} dz  \frac{(1-z)^N}{(1-z-u)^N}\prod_{i=1}^N
\frac{z-a_i+u}{z-a_i},
\eeq 
where we can already set all $a_i=0$ thus obtaining the formula for the Laguerre mean.

\begin{proposition}\label{prop-1}
The exponential mean for the Laguerre ensemble is given by the following exact formula:
\bea
\Bigl\langle \tr e^{uH} \Bigr\rangle=\frac 1u \frac 1{2\pi i} \oint_{C_0} dz  \frac{(1-z)^N}{(z+u-1)^N}
\frac{(z+u)^N}{z^N}&=&\frac 1u \frac 1{2\pi i} \oint_{C_0} dz  \Bigl(1+\frac 1{z+u-1}\Bigr)^N \Bigl(1-\frac 1z\Bigr)^N\nonumber\\
&:=&\frac 1u f_{N,N}(u)
\label{rep-N}
\eea
\end{proposition}

In the case of $s$ traces we have the corresponding $s$-fold integral over the variables $z_i$, $i=1,\dots,s$. The form of this integral is similar to the one in the Gaussian case (see \cite{BreHik}, \cite{MS10}), the main difference is that instead of adding exponentials we have to insert ratios $\frac{(1-z_i)^N}{(1-z_i+u_i)^N}$. We treat in details only the case $s=2$; the general idea will be clear from it.
\bea
\allowdisplaybreaks
&{}&
\Bigl\langle \tr e^{u_1H}\tr e^{u_2H}\Bigr\rangle^{\text{conn}}_a
=\sum_{k_1\ne k_2}\prod_{i=1}^N \frac1{(1-a_i)^N} \frac{(1-a_{k_1})^N}{(1-a_{k_1}-u_1)^N}\frac{(1-a_{k_2})^N}{(1-a_{k_2}-u_2)^N}\times\nonumber\\
&{}&\quad\times \Bigl[ \prod_{j\notin\{k_1,k_2\}} \frac{a_{k_1}+u_1-a_j}{a_{k_1}-a_j}\frac{a_{k_2}+u_2-a_j}{a_{k_2}-a_j}\Bigr]
\frac{a_{k_1}+u_1-a_{k_2}-u_2}{a_{k_1}-a_{k_2}} \nonumber\\
&{}&\qquad\qquad+ \frac{1}{u_1+u_2}f_{N,N}(u_1+u_2)-\frac 1{u_1u_2} f_{N,N}(u_1)f_{N,N}(u_2)
\nonumber\\
\allowdisplaybreaks
&{}&=\frac1{(2\pi i)^2}\oint_{C_0}\frac{dz_1}{u_1}\oint_{C_0}\frac{dz_2}{u_2}\Bigl[\prod_{j=1}^N\frac 
{(z_1+u_1-a_j)(z_2+u_2-a_j)}{(z_1-a_j)(z_2-a_j)}\Bigr] \frac{(1-z_1)^N (1-z_2)^N}{(z_1+u_1-1)^N(z_2+u_2-1)^N}\times\nonumber\\
&{}&\qquad\qquad\times \Bigl[ \frac{(z_1+u_1-z_2-u_2)(z_1-z_2)}{(z_1-z_2+u_1)(z_1-z_2-u_2)}-1\Bigr] +\frac{1}{u_1+u_2}f_{N,N}(u_1+u_2)\nonumber\\
&{}&\stackrel{a_j=0}{=} \frac1{(2\pi i)^2}\oint_{C_0}dz_1\oint_{C_0}dz_2\,\Bigl(1-\frac{1}{z_1}\Bigr)^N\Bigl(1-\frac{1}{z_2}\Bigr)^N
\Bigl(1+\frac{1}{z_1+u_1-1}\Bigr)^N \Bigl(1+\frac{1}{z_2+u_2-1}\Bigr)^N \times\nonumber\\
&{}&\qquad\qquad\times\frac{1}{(z_1-z_2+u_1)(z_1-z_2-u_2)} +\frac{1}{u_1+u_2}f_{N,N}(u_1+u_2).
\label{W2}
\eea
\begin{proposition}\label{prop-2}
Using the same replica method we can derive the integral representation for the one-loop mean in the case of generalized Laguerre ensemble corresponding to the Pastur--Marchenko law. For the rectangular $(\gamma_1N\times \gamma_2 N)$-matrices $B$ with $\gamma_2\ge \gamma_1$, we have
\beq
\frac{1}{\mathcal Z}\int DB\,DB^\dagger\,e^{-N\tr B B^\dagger} \tr e^{uBB^\dagger}
=\frac 1u \frac 1{2\pi i} \oint_{C_0} dz  \frac{(1-z)^{\gamma_2 N}}{(z+u-1)^{\gamma_2 N}}
\frac{(z+u)^{\gamma_1 N}}{z^{\gamma_1 N}}.
\label{MP-1loop}
\eeq
\end{proposition}

\section{The Harer--Zagier--Do--Norbury recursion equation}\label{s:lemma-u}
From now on, we concentrate on the case of the one-loop mean for the classic Laguere--Poisson ensemble.
\begin{lemma}\label{lem:DN}
[\cite{DN}] The function $f_{N,N}(u)$ in (\ref{rep-N}) satisfies the differential equation
\beq
\frac{\partial^2 f_{N,N}(u)}{\partial u^2} +\frac{4N}{u^2-1}\frac{\partial f_{N,N}(u)}{\partial u}-\frac{2N}{u(u^2-1)}f_{N,N}(u)=0.
\eeq
\end{lemma}
The {\bf proof} seems to be unexpectedly long for such a simple statement. It is nevertheless useful because it contains ingredients necessary when deriving differential equations for the one-loop mean in the generalized Laguerre ensemble case in Sec.~\ref{s:generalized}.

First, we introduce the functions 
\beq
f_{A,B}(u):= \frac 1{2\pi i} \oint_{C_0} dz  \Bigl(1+\frac 1{u+z-1}\Bigr)^A \Bigl(1-\frac 1z\Bigr)^B, \quad A,B\in {\mathbb Z}_{+}.
\label{fAB}
\eeq
Note that these functions enjoy the symmetricity relation obtained by swapping evaluation of the residue from $z=0$ to the pole at $z=1-u$. Accounting for the residue at the infinity, we obtain
\beq
f_{A,B}(u)=f_{B,A}(u) +(A-B)
\label{feat-1}
\eeq
Next, we can evaluate the derivative w.r.t. $u$. Using the identities
\bea
\frac{\partial f_{A,B}(u)}{\partial u}&=&
\frac 1{2\pi i} \oint_{C_0} dz  \biggl(\frac{\partial }{\partial u}\Bigl(1+\frac 1{u+z-1}\Bigr)^A\biggr) \Bigl(1-\frac 1z\Bigr)^B\nonumber\\
&=&\frac 1{2\pi i} \oint_{C_0} dz  \biggl(\frac{\partial }{\partial z}\Bigl(1+\frac 1{u+z-1}\Bigr)^A\biggr) \Bigl(1-\frac 1z\Bigr)^B\nonumber\\
&=&-\frac 1{2\pi i} \oint_{C_0} dz \Bigl(1+\frac 1{u+z-1}\Bigr)^A  \biggl(\frac{\partial }{\partial z}\Bigl(1-\frac 1z\Bigr)^B\biggr),\nonumber
\eea
we obtain (for the brevity, we omit the argument of the function $f_{A,B}(u)$)
\beq
\frac{\partial f_{A,B}}{\partial u}=-A(f_{A{-}1,B}-2f_{A,B}+f_{A{+}1,B})=-B(f_{A,B{-}1}-2f_{A,B}+f_{A,B{+}1})
\label{fAB-der}
\eeq
and the convenient representation for the second derivative reads:
\bea
\frac{\partial^2 f_{A,B}}{\partial u^2}&=&AB\bigl[ f_{A{-}1,B{-}1}+f_{A{-}1,B{+}1}+f_{A{+}1,B{-}1}+f_{A{+}1,B{+}1}-2f_{A{-}1,B}\bigr.\nonumber\\
&{}&\quad\bigl.-2f_{A,B{-}1}
-2f_{A{+}1,B}-2f_{A,B{+}1}+4f_{A,B}\bigr].
\label{fAB-der-der}
\eea
Using that
$$
\Bigl(1+\frac 1{u+z-1}\Bigr) \Bigl(1-\frac 1z\Bigr)=\frac{u}{u-1}\Bigl[ \Bigl(1+\frac 1{u+z-1}\Bigr) +\Bigl(1-\frac 1z\Bigr)\Bigr]-\frac{u+1}{u-1}\cdot 1
$$
we obtain the identity
\beq
f_{A,B}+\frac{u+1}{u-1} f_{A{-}1,B{-}1}-\frac{u}{u-1}\bigl[ f_{A{-}1,B}+f_{A,B{-}1}\bigr]=0, \quad A,B>0.
\label{fAB-quad}
\eeq
Using (\ref{fAB-quad}) we express all terms $f_{A{\pm}1,B{\pm}1}$ in (\ref{fAB-der-der}) via other terms in this relation obtaining
\bea
\frac{\partial^2 f_{A,B}}{\partial u^2}&=&AB\Bigl[ (f_{A{-}1,B}+f_{A,B{-}1})\Bigl( \frac{u}{u+1}+\frac{u+1}{u}-2\Bigr)
+(f_{A{+}1,B}+f_{A,B{+}1})\Bigl( \frac{u}{u-1}+\frac{u-1}{u}-2\Bigr)\Bigr.\nonumber\\
&{}&\qquad
-\Bigl.f_{A,B}\Bigl( \frac{u+1}{u-1}+\frac{u-1}{u+1}-2\Bigr)
\Bigr]\nonumber\\
&=&AB\Bigl[ (f_{A{-}1,B}+f_{A,B{-}1}) \frac{1}{u(u+1)}
+(f_{A{+}1,B}+f_{A,B{+}1}) \frac{1}{u(u-1)}\Bigr.\nonumber\\
&{}&\qquad \Bigl.-f_{A,B} \frac{4}{(u+1)(u-1)}\Bigr]
\label{fAB-der-der-1}
\eea
We can further simplify this expression if we set $A=B=N$. We can then use (\ref{feat-1})
to turn the right-hand side into a three-term expression; the result reads
\beq
\frac{\partial^2 f_{N,N}}{\partial u^2}=N^2 \Bigl[ (2f_{N,N{-}1}-1) \frac{1}{u(u+1)}
+(2f_{N,N{+}1}+1) \frac{1}{u(u-1)}-f_{N,N} \frac{4}{(u+1)(u-1)}\Bigr].
\label{der-3}
\eeq
 We need one more relation, which we obtain by integrating by parts in (\ref{fAB}):
 \begin{align}
 \frac 1{2\pi i} \oint_{C_0} dz  \Bigl(1+\frac 1{u+z-1}\Bigr)^A \Bigl(1-\frac 1z\Bigr)^B
 =&-\frac 1{2\pi i} \oint_{C_0} dz  \Bigl(1+\frac 1{u+z-1}\Bigr)^A\cdot B z\Bigl(1-\frac 1z\Bigr)^{B-1}\frac {1}{z^2}
 \nonumber\\
 &-\frac 1{2\pi i} \oint_{C_0} dz  \Bigl[\frac{\partial}{\partial z}\Bigl(1+\frac 1{u+z-1}\Bigr)^A\Bigr]\cdot z \Bigl(1-\frac 1z\Bigr)^B
 \nonumber\\
 =&\ \frac 1{2\pi i} \oint_{C_0} dz  \Bigl(1+\frac 1{u+z-1}\Bigr)^A\cdot B \Bigl[\Bigl (1-\frac 1z\Bigr)^{B} -(1-\frac 1z\Bigr)^{B-1}\Bigr]
 \nonumber\\
 &-\frac{\partial}{\partial u} \frac 1{2\pi i} \oint_{C_0} dz  \Bigl(1+\frac 1{u+z-1}\Bigr)^A\cdot z \Bigl(1-\frac 1z\Bigr)^B
 \label{IBP}
 \end{align}
 In the last line we have a new integral containing extra $z$ in the integrand. However, in the case $A=B=N$ we have the identity
 \beq
 \frac 1{2\pi i} \oint_{C_0} dz  \Bigl(1+\frac 1{u+z-1}\Bigr)^N \Bigl(1-\frac 1z\Bigr)^N (2z+u-1)=-uN,
 \label{id}
 \eeq 
 which follows from that the integrand is skew-symmetric w.r.t. the change of the integration variable $z\to -z-u+1$. So, performing this change of variables subsequently transferring the residue from $1-u$ to $0$, we remain only with the contribution from the residue at infinity, which is the right-hand side of the identity (\ref{id}).  We therefore have that
 $$
 \frac 1{2\pi i} \oint_{C_0} dz  \Bigl(1+\frac 1{u+z-1}\Bigr)^N \Bigl(1-\frac 1z\Bigr)^N z =\frac{1-u}{2} f_{N,N}(u)-\frac{uN}{2}
 $$
 and substituting it into (\ref{IBP}) we obtain the last required identity
 $$
 f_{N,N}=N[f_{N,N}-f_{N,N{-}1}]-\frac{\partial }{\partial u}\Bigl[\frac{1-u}{2}f_{N,N}-\frac{uN}{2}\Bigr]
 $$
 or
 \beq
 N[f_{N,N}-f_{N,N{-}1}] =\frac{(1-u)}{2}\frac{\partial f_{N,N}}{\partial u} +\frac {f_{N,N}}{2}-\frac{N}{2}.
 \label{feat-2}
 \eeq
 Using now (\ref{fAB-der}) for $A=B=N$ and relation (\ref{feat-2}) we can express $f_{N,N{-}1}$ and $f_{N,N{+1}}$ in terms of $f_{N,N}$ and its
 derivative $\partial f_{N,N}/\partial u$. Substituting the thus obtained expressions into (\ref{der-3}) we finally obtain the statement of the lemma.$\quad\square$

It is now technically straightforward to obtain the Do--Norbury three-term recursion relation. For this, we first introduce expansions of the one-loop means. We have that
\beq
W_1^{(g)}(x)=\sum_{n=0}^\infty C_n^{(g)} x^{-1-2g-n}
\label{Cgn}
\eeq
We then have the theorem
\begin{theorem}[\cite{DN}]
The coefficients $C_n^{(g)}$ of expansion of the one-loop means (\ref{Cgn}) satisfy the three-term recursion relation
\beq
(n+2g+1)C_n^{(g)}=(n+2g-2)(n+2g-1)^2 C_n^{(g-1)}+2(2n+4g-1)C_{n-1}^{(g)},\quad n>0,\ g>0.
\label{3-t}
\eeq
\end{theorem}

In order to prove this statement, note first that 
\beq
W'_1(x)=\frac{1}{-N}\int_0^\infty e^{-Nux}f_{N,N}(u).
\eeq
From Lemma~\ref{lem:DN} we have that
\beq
\int_0^\infty e^{-Nux} \Bigl[ u(u^2-1)\frac{\partial^2}{\partial u^2}f_{N,N}(u) +4Nu \frac{\partial}{\partial u}f_{N,N}(u) -2N f_{N,N}(u)
\Bigr] du=0
\eeq
Integrating by parts with accounting for the condition $f_{N,N}(0)=0$, we obtain 
\beq
\int_0^\infty e^{-Nux} f_{N,N}(u)\bigl[ N^2x^2 (u^3-u)-2Nx(3u^2-1)+6u +4N^2xu-4N-2N\bigr] du=0
\label{int-1}
\eeq
and since every factor of $u$ under the integral sign translates into $-\frac 1N \frac{\partial}{\partial x}$ outside the integral sign, combining all factors in
(\ref{int-1}), we obtain the differential equation on $W_1'(x)$:
\beq
-\frac 1N \Bigl[ x^2\frac{\partial^3}{\partial x^3} +6x \frac{\partial^2}{\partial x^2}+6 \frac{\partial}{\partial x}\Bigr] W_1'(x)
+N \Bigl[ x^2\frac{\partial}{\partial x} + 2x -4x \frac{\partial}{\partial x}-6\Bigr] W_1'(x)=0.
\label{int-2}
\eeq
Now, because $W_1'(x)=-\sum_{g=0}^\infty \sum_{n=0}^\infty N^{-2g} C_n^{(g)}(n+2g+1)x^{-2-2g-n}$, picking up terms of $N^{-2g+1}x^{-1-2g-n}$, we obtain
$$
-(n+2g)(n+2g-1)^2(n+2g-2)C_n^{(g-1)}+(n+1+2g)(n+2g)C_n^{(g)}-4(n+2g)(n+2g-1/2) C_{n-1}^{(g)}=0,
$$
which implies that these relations are empty at $n+2g=0$ and for $n+2g>0$ we obtain the three-term recursion relation (\ref{3-t}).

{\sl Case $g=0$}. For $g=0$ we have the differential equation 
$$
x(x-4)\frac{\partial^2}{\partial x^2}W_1^{(0)}(x)+2(x-3) \frac{\partial}{\partial x}W_1^{(0)}(x)=0,
$$
so
$$
\frac{\partial}{\partial x}W_1^{(0)}(x)=\frac{C}{x^{3/2}(x-4)^{1/2}},
$$
or
\beq
W_1^{(0)}(x)=-\frac{t_0}{2}\frac{\sqrt{x-4}}{\sqrt{x}}+\frac{t_0}{2},
\eeq
where $t_0$ is the (normalized) number of eigenvalues of the matrix $H$, which we set equal to the unity in what follows, $t_0=1$. 

{\sl Cases $n=0$ and $n=1$.} For the lowest expansion terms we have: $C_0^{(0)}$ is arbitrary. The recursion for $n=0$ reads
$$
-(2g-1)^2(2g-2) C_0^{(g-1)}+(2g+1)C_0^{(g)}=0,\quad g\ge1,
$$
so for $g=1$ we have that $3C_0^{(1)}=0$, so this coefficient as well as all other $C_0^{(g)}$ vanish,
\beq
C_0^{(g)}=0,\ g\ge1, \quad C_0^{(0)}=1.
\label{C0g}
\eeq
So, for $n=1$ with accounting for (\ref{C0g}), we obtain
\beq
-(2g)^2(2g-1)C_1^{(g-1)}+2(g+1) C_1^{(g)}-2(4g+1)\delta_{g,0}=0,\quad g\ge 0,
\eeq
which implies that $C_1^{(0)}=1$ and
\beq
C_1^{(g)}=\frac{2^g g! (2g-1)!!}{(g+1)}.
\label{C1g}
\eeq
As expected, all $C_1^{(g)}$ are positive integers. In principle, because $C_k^{(g)}$ enumerate graphs with trivial groups of automorphisms, we
expect all of them to be positive integers.

\section{Recursion relations in terms of variables $v_{k}(\lambda)$}\label{s:sml}

The three-term recursion relation (\ref{3-t}) looks rather attractive. However, for a given genus $g$ we still have infinitely many coefficients $C_k^{(g)}$ to be determined. On the other hand, we know from the topological recursion and from our experience with the Gaussian model in \cite{ACNP2} that at a given genus $g$ of the $1/N$-expansion, actual $s$-point correlation functions (in the stable cases $2g-2+s>0$), at least for spectral curves associated with ensembles of classic orthogonal polynomials (Gaussian, Laguerre, and Legendre), are finite linear combinations of products of the functions
\beq
s_{k, \beta}(\lambda):=\frac{(e^\lambda+e^{-\lambda})^\beta}{(e^\lambda-e^{-\lambda})^{2k+3}},\quad 0\le k\le 3g+s,\ \beta=0,1.
\label{skb}
\eeq 
that is 
\beq
W_s^{(g)}(x_1,\dots, x_s)dx_1\cdots dx_s
=\sum_{\{k_i,\beta_i\}} C_{\{k_i,\beta_i\}} \prod_{i=1}^s [s_{k_i,\beta_i}(\lambda_i)]dx_i,\quad x_i=e^{\lambda_i}+e^{-\lambda_i}+2,
\label{Wsg}
\eeq
with restrictions (\ref{skb}) on the range of summation w.r.t. $k_i$. So, any $W_s^{(g)}$ is completely defined by a finite set of (presumably integer)
coefficients $C_{\{k_i,\beta_i\}}$. All these coefficients can be found, order by order in $g$ and $s$, starting from two basic functions $W_3^{(0)}$ and
$W_1^{(1)}$ using the topological recursion procedure of \cite{Ey}, \cite{ChEy}, \cite{CEO}, \cite{EOrTop}
based on spectral curves of the corresponding ensembles: $y^2-xy+1=0$ (Gaussian), $xy^2-x+4=0$ (Laguerre), and $y^2(x^2-4)-1=0$ (Legendre). 

In the Laguerre ensemble case, we have that
\beq
W_3^{(0)}(x_1,x_2,x_3)dx_1dx_2dx_3=\prod_{i=1}^3 \Bigl[ (s_{0,1}(\lambda_i)+2s_{0,0}(\lambda_i)dx_i\Bigr]
\label{W30}
\eeq
and
\beq
W_1^{(1)}(x)dx=[s_{1,1}(\lambda)+2s_{1,0}(\lambda)]dx.
\label{W11}
\eeq
These formulas can be obtained straightforwardly from the input of the Laguerre ensemble: 
we have a two-sheet covering of the sphere with coordinates $p:=e^\lambda$ and $\bar p:=e^{-\lambda}$ and with the standard involution $\lambda\to -\lambda$. The relevant functions are
$x=e^\lambda+e^{-\lambda}+2$ and $y=(e^\lambda-1)/(e^\lambda+1)$, $ydx=e^{-\lambda}(e^\lambda-1)^2d\lambda$,
the recursion kernel (a $(1,-1)$-differential) is 
$$
K(\mu,\lambda):=dE(\mu,\lambda) \frac1{ydx}=\frac{de^\mu}{e^\mu-e^\lambda}\frac{e^\lambda}{(e^\lambda-1)^2d\lambda},
$$
and the universal Bergmann kernel (a symmetric bi-differential) reads
$$
B(p,q)=\frac{dp\,dq}{(p-q)^2}=\frac{de^\lambda\, de^\mu}{(e^\lambda-e^\mu)^2}
$$
for any model on a sphere. 
All $s$-point correlation functions are interpreted as symmetric $s$-differentials;
the two correlation functions initiating the recursion procedure are
$$
W_3^{(0)}(x_1,x_2,x_3)=\sum_{\lambda=0,i\pi}\mathop{\hbox{res}}_\lambda 
K(p_1,\lambda)(B(p_2,\lambda) + B(\bar p_2,\lambda))(B(p_3,\lambda) + B(\bar p_3,\lambda))
$$
(here actually only the residue at $\lambda=0$ is nonzero) and
$$
W_1^{(1)}(x_1)=\sum_{\lambda=0,i\pi}\mathop{\hbox{res}}_\lambda K(p_1,\lambda) B(\lambda,-\lambda).
$$
Evaluating these integrals we obtain the respective expressions (\ref{W30}) and (\ref{W11}) above. 

However, in the Laguerre one-loop mean case, recursion relations take a more instructive form when being expressed in terms of \emph{different} times, namely
\beq
v_k(\lambda)=\Bigl[\frac{e^{\lambda}-1}{e^{\lambda}+1}\Bigr]^{2k+1},\quad k\in {\mathbb Z}.
\label{vk}
\eeq
We first transform the differential operators in (\ref{int-2}): the first one is
$$
\Bigl[x^2\frac{\partial}{\partial x} + 2x -4x \frac{\partial}{\partial x}-6\Bigr]\frac{\partial}{\partial x}W_1^{(g-1)}=\frac{\partial}{\partial x}
\Bigl[x(x-4)\frac{\partial}{\partial x}-2\Bigr]W_1^{(g-1)}=\frac{\partial}{\partial x} \Bigl[ (e^\lambda-e^{-\lambda})\frac{\partial}{\partial \lambda}-2\Bigr]W_1^{(g-1)}
$$
and the second operator reads
$$
\Bigl[x^2\frac{\partial^4}{\partial x^4} +6x \frac{\partial^3}{\partial x^3}+6 \frac{\partial^2}{\partial x^2}\Bigr]W_1^{(g)}=
\frac{\partial}{\partial x}\Bigl[ \frac{\partial}{\partial x} \Bigl(x \frac{\partial}{\partial x}+1\Bigr) x \frac{\partial}{\partial x}\Bigr]W_1^{(g)},
$$
so we can disregard the leftmost differentiation operator $\frac{\partial}{\partial x}$ and obtain the recursion formula for $W_1^{(g)}(\lambda)$:
\beq
\Bigl[ (e^\lambda-e^{-\lambda})\frac{\partial}{\partial \lambda}-2\Bigr] W_1^{(g-1)}(\lambda)
=\frac{\partial}{\partial x} \Bigl(x \frac{\partial}{\partial x}+1\Bigr) x \frac{\partial}{\partial x} W_1^{(g)}(\lambda)
\label{rec-v1}
\eeq
Now it is clear why we replace the standard variables $s_{k, \beta}(\lambda)$ by $v_k(\lambda)$: the action of the first operator in (\ref{rec-v1}) on $v_k(\lambda)$ is merely
$$
\Bigl[ (e^\lambda-e^{-\lambda})\frac{\partial}{\partial \lambda}-2\Bigr]v_k(\lambda)=4k v_k(\lambda).
$$
We need now the action of the second operator in (\ref{rec-v2}) on $v_k$. for this, note that 
$$
x \frac{\partial}{\partial x} v_k(\lambda)=\frac{2k+1}{2}\bigl[ v_{k-1}(\lambda)-v_k(\lambda) \bigr],\qquad
\frac{\partial}{\partial x} v_k(\lambda)=\frac{2k+1}{8}\bigl[ v_{k-1}(\lambda)-2v_k(\lambda) +v_{k+1}(\lambda)\bigr],
$$
so, putting everything together, we obtain
\begin{align*}
\partial_x(x\partial_x+1) x\partial_x v_k=&\ \frac{(2k+1)(2k-1)(2k-3)}{32}v_{k-3}- \frac{(2k+1)(2k-1)(k-1)}{4}v_{k-2}\\
&+\frac{3(2k+1)(2k-1)^2}{16}v_{k-1}-\frac{(2k+1)k(2k-1)}{4}v_{k}+\frac{(2k+1)^2(2k-1)}{32}v_{k+1},
\end{align*}
and we come to the following lemma.
\begin{lemma}\label{lem:vk}
The genus filtrated one-loop means $W_1^{(g)}(\lambda)$ are given by finite expansions in the variables $v_k(\lambda)$ (\ref{vk}),
$$
W_1^{(g)}(\lambda)=\sum_{k=-3g}^{g}a_k^{(g)} v_k(\lambda),
$$
where the coefficients $a_k^{(g)}$ are determined recursively: for $k\ne 0$, we have
\begin{align}
4k a_k^{(g)}=&\ \frac{(2k+7)(2k+5)(2k+3)}{32}a_{k+3}^{(g-1)}-\frac{(2k+5)(2k+3)(k+1)}{4}a_{k+2}^{(g-1)}\nonumber\\
&+\frac{3(2k+3)(2k+1)^2}{16}a_{k+1}^{(g-1)}-\frac{(2k+1)k(2k-1)}{4}a_{k}^{(g-1)}+\frac{(2k-1)^2(2k-3)}{32}a_{k-1}^{(g-1)},
\label{rec-v2}
\end{align}
and we determine the coefficient $a_0^{(g)}$ from the asymptotic condition $W_1^{(g)}(\lambda)\to 0$ as $\lambda\to +\infty$, which implies that
\beq
\sum_{k=-3g}^{g} a_k^{(g)}=0
\label{asymptotic}
\eeq
for $g>0$. The starting term of the recursion is
$$
a_k^{(0)}=-\delta_{k,0}/2.
$$
\end{lemma}
Recurrent relations  (\ref{rec-v2}) are consistent provided they are satisfied at $k=0$, which implies that 
\beq
\frac{105}{32} a_{3}^{(g)} -\frac{15}{4} a_{2}^{(g)} +\frac{9}{16} a_{1}^{(g)}-\frac{3}{32} a_{-1}^{(g)}=0\quad \forall g.
\label{consistency}
\eeq
The consistency of recurrent relations (\ref{rec-v2}) follows from the fact of existence of quantities $W_1^{(g)}(\lambda)$ determined by the standard topological recursion procedure. The only singularities of the
obtained 1-differentials $W_1^{(g)}(x)dx$ are poles of degrees not higher than $6g-2$ at $x=4$, or at $e^\lambda=1$, and not higher than $g-1$ at $x=0$, or at $e^\lambda=-1$. Together with the asymptotic condition $W_1^{(x)}\sim 1/x+O(x^{-2})$ as
$x\to \infty$ and the skew-symmetricity of $W_1^{(g)}(\lambda)$ w.r.t. the change of variables $\lambda\to -\lambda$ it implies that we can always express $W_1^{(g)}(\lambda)$ in terms of $v_k(\lambda)$.

The asymptotic conditions (\ref{Cgn}) and (\ref{C0g}) imply that the first $2g+2$ moments of the coefficients $a_k^{(g)}$ vanish,
\beq
\sum_{k=-3g}^{g} k^r a_k^{(g)}=0,\quad k=0,1,\dots, 2g+1,\quad g\ge 1.
\label{asym-r}
\eeq

As an example, we find the first two terms: $W_1^{(1)}(\lambda)$ and $W^{(2)}_1(\lambda)$. Having 
$W_1^{(0)}(\lambda)=-\frac12 v_0(\lambda)+\frac12$, we obtain for nonzero coefficients of $W^{(1)}_1$ and $W^{(2)}_1$:
\bea
&{}&
a_{r}^{(1)}=\Bigl._{r=-3}\Bigl\{ \frac{1}{2^{8}},\ \frac{-4}{2^{8}},\ \frac{6}{2^{8}}, \ \frac{-4}{2^{8}}, \ \frac{1}{2^{8}}\Bigr\}_{r=1}\Bigr.;
\nonumber\\
&{}&
a_{r}^{(2)}=\Bigl._{r=-6}\Bigl\{\frac{105}{2^{16}}, \ \frac{-616}{2^{16}}, \ \frac{1500}{2^{16}}, \ \frac{-1944}{2^{16}},
\ \frac{1430}{2^{16}}, \ \frac{-600}{2^{16}}, \ \frac{156}{2^{16}}, \ \frac{-40}{2^{16}},
\ \frac{9}{2^{16}}\Bigr\}_{r=2}\Bigr..\nonumber
\eea
In these expressions, we determine the coefficients $a_0^{(1)}$ and $a_0^{(2)}$ from the asymptotic condition (\ref{asymptotic}). One can also verify that both the consistency condition (\ref{consistency}) and the asymptotic conditions (\ref{asym-r}) hold in the both expressions. In every term of the genus expansion, the one-loop correlation function $W_1^{(g)}$ is therefore completely determined by $4g+1$ coefficients $a_k^{(g)}$, ($-3g\le k\le g$) subject to two linear constraints (\ref{asymptotic}) and (\ref{consistency}) holding independently of the expansion order and to $2g+1$ additional moment constraints (\ref{asym-r}) (for $g\ge1$) following from the asymptotic behaviour of $W_1^{(g)}(x)$.

\section{Recurrent relations for a generalized Laguerre ensemble}\label{s:generalized}

In this section, we first present details of derivation of new Harier--Zagier-like recurrent relations for the first nontrivial case of the generalized Laguerre ensemble (\ref{BB-1}) with $\gamma_2 N=\gamma_1 N+1=N+1$. As we demonstrate below, in this case we obtain a fourth-order differential equation on the one-loop mean obtained from a second-order differential equation on $\bigl\langle \tr e^{uH} \bigr\rangle$. We then describe a general method of deriving a differential equation on $\bigl\langle \tr e^{uH} \bigr\rangle$ for any $\gamma_2 N=N+k$; this differential equation is of order $2k$ and  it can be translated into a differential equation on the standard one-loop mean. We present the corresponding $4$th order linear differential equation for $k=2$.

We begin with transforming the integral (\ref{MP-1loop}) into a function $f_{A,B}$: for $\gamma_1 N=N$ and $\gamma_2 N=N+k$, using the substitution $z=-uz'$, we obtain
\bea
\frac 1u \frac 1{2\pi i} \oint_{C_0} dz  \frac{(1-z)^{N+k}}{(z+u-1)^{N+k}}
\frac{(z+u)^{N}}{z^{N}}
&=&(-1)^{N+k-1} \frac 1{2\pi i} \oint_{C_0} dz'  \Bigl(1+\frac{1}{z'+1/u-1}\Bigr)^{N+k}\Bigl(1-\frac{1}{z'}\Bigr)^{N}\nonumber\\
&=&(-1)^{N+k-1} f_{N+k,N}(1/u),\label{T-1}
\eea
so finding a differential equation on $f_{N+k,N}(x)$ for fixed $N$ and $k$ is the crucial step of the construction. 

{\bf The case $k=1$.} We present details of calculation only in the case $k=1$. There, we begin with expressing
$f_{N,N}$ and $f_{N+1,N+1}$ through $f_{N+1,N}$ and $\partial_u f_{N+1,N}$. For this, we first substitute (\ref{fAB-der}) for the derivative
of $f_{N,N}$ in the relation (\ref{feat-2}) obtaining the relation
$$
\frac{N(u-1)}{2}f_{N,N+1}-uN f_{N,N}+\frac{N(u+1)}{2}f_{N,N-1}+\frac{f_{N,N}}{2}-\frac{N}{2}=0,
$$
or, taking into account that $f_{N,N+1}=f_{N+1,N}-1$ (see (\ref{feat-1})) and applying (\ref{fAB-quad}) at $(A,B)=(N+1,N)$ we obtain a simple
equation
$$
-uN f_{N,N}+uN f_{N+1,N-1}+f_{N,N}-uN=0
$$
in which we express $f_{N+1,N-1}$ using again (\ref{fAB-der}) for the derivative of $f_{N+1,N}$ obtaining an equation that contains $f_{N,N}$, $f_{N+1,N+1}$, $f_{N+1,N}$, and $\partial_u f_{N+1,N}$:
\beq
(-uN+1)f_{N,N}-uN+u\Bigl(-\frac{\partial f_{N+1,N}}{\partial u} +2N f_{N+1,N}-N f_{N+1,N+1}\Bigr)=0
\label{T-2}
\eeq
Another constraint on $f_{N+1,N+1}$, $f_{N,N}$, and $f_{N+1,N}$ follows from (\ref{fAB-quad}) for $(A,B)=(N+1,N+1)$ with accounting for
that $f_{N,N+1}=f_{N+1,N}-1$:
\beq
(u+1)f_{N,N}-2u f_{N+1,N}+(u-1)f_{N+1,N+1}+u=0.
\label{T-3}
\eeq
Using (\ref{T-2}) and (\ref{T-3}) we obtain
\beq
\Bigl( \frac{uN}{u-1}+\frac 12\Bigr)f_{N,N}=\frac u2 \frac{\partial f_{N+1,N}}{\partial u}+\frac{uN}{u-1}f_{N+1,N}-\frac{uN}{2(u-1)}.
\label{T-4}
\eeq
From (\ref{fAB-der-der-1}), we have that
$$
\frac{\partial^2 f_{N+1,N}}{\partial u^2}=N(N+1)\Bigl[ \frac{f_{N,N}+f_{N+1,N-1}}{u(u+1)} +\frac{f_{N+2,N}+f_{N+1,N+1}}{u(u-1)}
-\frac{4f_{N+1,N}}{u^2-1}\Bigr],
$$
where we can now express all terms in the r.h.s. (recall that $(N+1)f_{N+2,N}=-\partial_u f_{N+1,N}-(N+1)f_{N,N}+2(N+1) f_{N+1,N}$)
in terms of $f_{N+1,N}$ and $\partial_u f_{N+1,N}$. After some straightforward algebra, we come to the lemma
\begin{lemma}\label{lem:fN+1N}
The function $f_{N+1,N}(u)$ satisfies the second-order differential equation
\bea
\Bigl(\frac{uN}{u-1}+\frac 12\Bigr)\frac{\partial^2 f_{N+1,N}}{\partial u^2}&=&
\frac{-8u^2N^2 +(-8u^2+4u)N -(u-1)^2}{2u(u^2-1)(u-1)}\frac{\partial f_{N+1,N}}{\partial u}\nonumber\\
&{}&\quad +\frac{2N(N+1)}{(u^2-1)(u-1)}f_{N+1,N}-\frac{N(N+1)}{(u^2-1)(u-1)}.\label{T-5}
\eea
\end{lemma}

Using this lemma and the fact that the actual one-loop mean is related to $f_{N+1,N}$ by the integral formula
\beq
W_1(x)=\int_0^\infty e^{-Nux}f_{N+1,N}(1/u)du,
\eeq
substituting (\ref{T-5}) and integrating by parts, we finally come to the following corollary.
\begin{corollary}
The one-loop correlation function $W_1(x)$ for the generalized Laguerre ensemble with $\gamma_2N=\gamma_1N+1=N+1$ satisfies the following inhomogeneous differential equation:
\bea
&{}&\Bigl[ N^2 \Bigl((-2x^2+8x)\frac{\partial}{\partial x}+4\Bigr)+N \Bigl((-x^2+8x)\frac{\partial}{\partial x}+4\Bigr)\Bigl.\nonumber\\
&{}&\qquad+1\Bigl(2x^2 \frac{\partial^3}{\partial x^3} +(-x^2+12x)\frac{\partial^2}{\partial x^2}+(-x+12)\frac{\partial}{\partial x} +1\Bigr)
\nonumber\\
&{}&\qquad\qquad\Bigl.+N^{-1}\Bigl( x^2\frac{\partial^3}{\partial x^3}+6x\frac{\partial^2}{\partial x^2}+6\frac{\partial}{\partial x}\Bigr)
+N^{-2}\Bigl( x^2\frac{\partial^4}{\partial x^4}+7x\frac{\partial^3}{\partial x^3}+9\frac{\partial^2}{\partial x^2}\Bigr)\Bigr] W_1(x)\nonumber\\
&=&2N^2+N+\frac{2N+2}{x}
\label{W1}
\eea
\end{corollary}

The first terms of expansion of the solution to (\ref{W1}) up to the order $O(x^{-6})$
are in accordance with the diagrammatic representation for unicellular bicolored maps in which cycles of one color enter with the
multiplier $N$ and cycles of the other color enter with the multiplier $N+1$:
\bea
W_1(x)&=&\frac 1x+\Bigl(1+\frac 1N \Bigr)\frac1{x^2}+\Bigl(2+\frac 1N \Bigr)\Bigl(1+\frac 1N \Bigr)\frac1{x^3}
+\Bigl(5+\frac 5N+\frac{2}{N^2} \Bigr)\Bigl(1+\frac 1N \Bigr)\frac1{x^4}\nonumber\\
&{}&+\Bigl(7+\frac 7N+\frac{6}{N^2} \Bigr)\Bigl(2+\frac 1N \Bigr)\Bigl(1+\frac 1N \Bigr)\frac1{x^5}
+O(x^{-6})
\eea

\begin{corollary}
Introducing the $1/N$-expansion of $W_1(x)=\sum_{r=0,1/2,1,3/2,\dots} N^{-2r} W_1^{(r)}(x)$ (the terms of this expansion with
a half-integer index $r$ vanish in the case $k=0$) and expanding every term of this ``fractional genus expansion'' 
in a series in $x$,
$$
W_1^{(r)}(x)=\sum_{n=0}^\infty C_n^{(r)} x^{-1-s-2r},
$$  
we find that the coefficients $C_n^{(r)}$ of these series satisfy the eight-term inhomogeneous recursion relation:
\bea
&{}&(n+1)\bigl(2C^{(r)}_{n-2r}+C^{(r-1/2)}_{n-2r+1}\bigr)-4(2n-1)\bigl(C^{(r)}_{n-2r-1}+C^{(r-1/2)}_{n-2r}\bigr)\nonumber\\
&{}&-(n-1)(n-2)(n-3)\bigl(2C^{(r-1)}_{n-2r}+C^{(r-3/2)}_{n-2r+1}\bigr)-(n+1)(n-1)C^{(r-1)}_{n-2r+1}\nonumber\\
&{}&+(n-1)(n-2)(n-3)^2 C^{(r-2)}_{n-2r+1}=2\delta_{r,0}\delta_{n,0}+\delta_{r,1/2}\delta_{n,0}+2\delta_{r,1/2}\delta_{n,1}+2\delta_{r,1}\delta_{n,1}
\eea
valid at all $n$ and $r$ providing all $C^{(r)}_{n}=0$ for $r<0$ or/and $n<0$. 
\end{corollary}

{\bf The case of higher $k$.} For $k\ge 2$, relations (\ref{fAB-der}), (\ref{fAB-quad}), (\ref{fAB-der-der-1}), and (\ref{feat-2}) suffice to derive a closed inhomogeneous linear differential equation of the $2k$th order on $f_{N+k,N}$. For this, we first rewrite (\ref{T-2}) and (\ref{T-4}) as three-term relations
\begin{align}
&((N+1)u-1)((u+1)f_{N,N}-2uf_{N+1,N})+(N+1)u(u-1)f_{N+2,N}=-Nu,\label{T-5a}\\
&((N+1)u-1)((u-1)f_{N+2,N+2}-2uf_{N+2,N+1})+(N+1)u(u+1)f_{N+2,N}=(N+2)u.\label{T-5b}
\end{align}
The strategy of deriving the differential equation on $f_{N+k,N}$ is as follows: note first that using (\ref{fAB-der}) and (\ref{fAB-quad}) we 
can present an $n$th derivative of $f_{N+k,N}$ as a linear combination (with coefficients being functions of $u$ and $N$) of terms 
$f_{N+k-s,N},\dots, f_{N+k-1,N}, f_{N+k,N}, f_{N+k,N+1},\dots, f_{N+k,N+n-s}$ (for any $0<s<n$) and of derivatives of $f_{N+k,N}$ of orders lesser or equal $\max(s,n-s)$. The locus of participating $f_{A,B}$ has a hook structure in the discrete lattice of indices $(A,B)$ with the break at $(A,B)=(N+k,N)$ and legs of nonzero lengths $s$ and $n-s$ stretched towards the diagonal, i.e., the locus of $f_{A,A}$. We then present the derivative of order $2k$ as a hook $(N,N)$---$(N+k,N)$---$(N+k,N+k)$ and use linear constraints (\ref{T-5a}) and (\ref{T-5b}) to reduce it to the hook $(N,N+1)$---$(N+k,N)$---$(N+k,N+k-1)$. We represent the $(2k-1)$th derivative as a hook $(N,N+1)$---$(N+k,N)$---$(N+k,N+k)$ and use (\ref{T-5b}) to reduce it again to the hook $h_{2k-2}:=(N,N+1)$---$(N+k,N)$---$(N+k,N+k-1)$. We then represent all derivatives starting  from the second derivative to the derivative of order $2k-2$ using a sequence of strictly embedded hooks $h_2\subset h_3\subset \cdots \subset h_{2k-3}\subset h_{2k-2}$ thus obtaining a system of $2k-1$ algebraically independent (because the highest term of each equation is the derivative of the corresponding order) inhomogeneous equations linear in $f_{N+1,N},\dots, f_{N+k,N},\dots, f_{N+k,N+k-1}$ and in the derivatives of $f_{N+k,N}$ of lower orders. Using last $2k-2$ of these equations, we can express all $2k-2$ variables $f_{N+1,N},\dots, f_{N+k-1,N}$ and $f_{N+k,N+1},\dots f_{N+k,N+k-1}$ in terms of derivatives of $f_{N+k,N}$ up to the order $2k-1$ and the last remaining differential equation of order $2k$ then becomes the constraint differential equation on $f_{N+k,N}$ using which one can derive the recurrent relation on the corresponding terms of genus expansion of $W_1(x)$ containing a finite number of terms. 

We very briefly describe the result for $k=2$. After evaluating $f^{IV}:=\partial^4 f_{N+2,N}/\partial u^4$, $f''':=\partial^3 f_{N+2,N}/\partial u^3$,
$f'':=\partial^2 f_{N+2,N}/\partial u^2$ and denoting $f':=\partial^4 f_{N+2,N}/\partial u^4$, we obtain
\begin{align}
f^{IV}=&\ \frac{2N(N{+}1)(N{+}2)}{u(u^2-1)}\Bigl[ \frac{(4N{+}10)u^2+(8N{+}6)u-4(N{+}1)}{u^2(u^2-1)(u-1)}f_{N+2,N+1}\Bigr.\nonumber\\
&+\frac{(4N{-}2)u^2-(8N{+}10)u-4(N{+}1)}{u^2(u^2-1)(u+1)}f_{N+1,N}\nonumber\\
&-\frac{(8(N{+}1)^2+2)u^3+12(N{+}1)u^2 +(8(N{+}1)^2-2)u-4(N{+}1)}{u(u^2-1)^2((N{+}1)u-1)}f_{N+2,N}\nonumber\\
&+\Bigl.\frac{(4N^2{+}8N{+}6)(u^2+1)+12(N{+}1)u}{(u^2-1)^2((N{+}1)u-1)}\Bigr]
-\frac{6u^2+2(N{+}1)u-4}{u(u^2-1)}f''' -\frac{6u+4(N{+}1)}{u(u^2-1)}f'';\\
f'''=&-\frac{2N(N{+}1)(N{+}2)}{u(u^2-1)}\Bigl[  \frac{2}{u-1}f_{N+2,N+1}+\frac{2}{u+1}f_{N+1,N}-\frac{4(N{+}1)u^2}{(u^2-1)((N{+}1)u-1)}f_{N+2,N}
\Bigr.\nonumber\\
&\ \Bigl. +\frac{2u((N{+}1)u+1)}{(u^2-1)((N{+}1)u-1)}\Bigr] -\frac{3u^2+2(N{+}1)u-3}{u(u^2-1)}f'' -\frac{2(N{+}1)}{u(u^2-1)}f';\\
f''=&\ \frac{2N(N{+}2)}{u(u^2-1)}(f_{N+2,N+1}-f_{N+1,N})-\frac{2(N{+}1)u-2}{u(u^2-1)}f'.
\end{align}
Expressing $f_{N+2,N+1}$ and $f_{N+1,N}$ from the last two equations and substituting the result into the first one, we come to the following
linear differential equation for $f:=f_{N+2,N}$.
\begin{lemma}\label{lem:fN+2N}
The function $f:=f_{N+2,N}(u)$ satisfies the inhomogeneous fourth-order differential equation
\begin{align}
&f^{IV}+2\frac{3u^3+2(N{+}1)u^2-u+(N{+}1)}{u^2(u^2-1)}f'''\nonumber\\
&+2\frac{3u^5+2(N{+}1)u^4-2(N{+}1)^2u^3+3(N{+}1)u^2+(6(N{+}1)^2-3)u -3(N{+}1)}{u^3(u^2-1)^2}f'' \nonumber\\
& -2(N{+}1)\frac{4(N{+}1)u^2+8N(N{+}2)u +10(N{+}1)}{u^3(u^2-1)^2}f' 
+\frac{4N(N{+}1)(N{+}2)(u+2(N{+}1))}{u^2(u^2-1)^2((N{+}1)u-1)}(f-1)=0
\end{align}
\end{lemma}

\section{Conclusion}
It seems interesting to re-derive the present results using a variant of the reproducing kernel technique (see \cite{Mehta}) used in \cite{Ledoux} and \cite{WF} for deriving recurrent formulas of Harer--Zagier type for ensembles of orthogonal and symplectic matrices.

In the forthcoming paper \cite{ACN} we shall extend to the case of irregular spectral curves the topological recursion formulation in terms of a dynamics of Young diagrams developed in \cite{ACNP2} in application to the Gaussian ensemble. It is also tempting to probe the existence of a Givental-like decomposition (see \cite{Giv}) of form analogous to that in \cite{Ch95} for irregular spectral curves. Another interesting question is about the meaning of the consistency condition (\ref{consistency}) from the point of view of bicolored maps: indeed, if we rewrite this condition in terms of $x$-variables, we obtain the exact identity on the one-loop mean $W_1(x)dx$:
\beq
\frac1{2\pi i}\oint_{\mathcal C_0} dx\frac{3x^3+16x-16}{x^{3/2}(x-4)^{9/2}}\bigl[ W_1(x)-1/2\bigr]=0,
\eeq
which holds separately in every order of $1/N$-expansion.
It is straightforward to verify this identity, e.g., for the first two terms of the expansion: $W_1^{(0)}(x)-1/2=-\frac12 x^{1/2}(x-4)^{-1/2}$
and $W_1^{(1)}(x)=x^{-3/2}(x-4)^{-5/2}$.

It is also tempting to find a geometrical interpretation of generalized Laguerre ensembles. Shortly after the first variant of this text was sent to the arXiv, the text \cite{Safnuk} appeared in which the author conjectured that the Kontsevich--Penner models with higher $k$ ($Q$ in his notation) can be generating functions for $Q$-graded correlation functions for open intersection numbers. This indicates that matrix models with higher $k$ (or $Q$) certainly deserve further studies.

\end{document}